\documentclass[11pt,twoside]{article}
\usepackage{asp2010}

\resetcounters

\begin{document}

\title{Predicting the Evolutionary Descendents of Sequence E Stars}
\author{J. D. Nie$^{1,2}$, P. R. Wood$^1$, and C. P. Nicholls$^1$
\affil{$^1$Research School of Astronomy and Astrophysics, Australian National
University, Cotter Road, Weston Creek ACT 2611, Australia}}
\affil{$^2$Department of Astronomy, Beijing Normal University, Beijing, 100875, China}

\begin{abstract}
Sequence E variables are close binary red giants that show ellipsoidal
light variations. They are likely to terminate their red giant evolution 
by a common envelope (CE) event when the red giant fills its Roche lobe, 
and produce close binary Planetary Nebulae (PNe). We made a Monte Carlo 
simulation to predict the fraction of Planetary Nebulae Nuclei (PNNe) that
are post-CE binaries, using the observed frequency of sequence E binaries 
in the LMC to normalize our calculations. We find that 10-16\% of PNNe should  
be short period, post-CE binaries.
\end{abstract}

\section{Introduction}\label{sect1}
Sequence E variables are ellipsoidal binaries which follow a loose
period-luminosity relation \citep{1999IAUS..191..151W,2004AcA....54..347S}. 
In the LMC, they make up 0.5 to 2\% of luminous red giant stars. Most of 
the sequence E stars lie on the first giant branch in the case of low mass 
stars \citep{1999IAUS..191..151W}, and some of them also evolve to the 
AGB \citep{2004AcA....54..347S,2007ApJ...660.1486S}. While all the sequence 
E stars show ellipsoidal variations, about 7\% of them show eclipses in 
addition to the ellipsoidal light variations \citep{2004AcA....54..347S}. 
In the ellipsoidal binary systems, the primary star is a red giant 
substantially filling its Roche lobe, and the secondary star is an unseen 
companion orbiting the primary star. Due to the tidal interaction, the 
primary star is distorted by the companion, causing an ellipsoid or pear-like 
shape. As the system orbits around, the change in the apparent surface area 
gives rise to the ellipsoidal light variation. 

Sequence E stars are likely the precursors of close binary Planetary 
Nebulae (PNe). In the ellipsoidal binary systems, the red giant primary 
star has already substantially filled its Roche lobe. As the star evolves, 
it will experience a Roche lobe overflow and then faces a common envelope
(CE) event. The primary star will terminate its red giant phase by 
ejecting the CE and it will produce a close binary PN.

At the present time the most favoured theory explaining the nonspherical
PNe is the binary hypothesis \citep{1990ApJ...355..568B, 1993ApJ...418..794Y,
1997ApJS..112..487S, 2000ASPC..199..115B, 2007BaltA..16...79Z, 2009PASP..121..316D}.
Theoretical considerations suggest that a binary with  CE will eject the 
entire envelope to produce an asymmetrical PN, with elliptical or bipolar 
shape. In fact, PNe with close binary nuclei have been confirmed by observations
\citep{1987fbs..conf..221B,1992IAUS..151..517B,1994ASPC...56..179B,2000ASPC..199..115B,
2009A&A...496..813M}. However, about their fraction, little is known. 
Bond and collaborators found about 10-15\% of PNe with close binary central 
stars, while \citet{2009A&A...496..813M} found 12-21\% of  Planetary Nebulae Nuclei (PNNe)
are close post-CE binaries.

Sequence E stars are likely to produce close binary PNe, and their fraction 
is known. In the LMC, \citet{1999IAUS..191..151W} found 0.5\% of the red 
giants on the top 1 magnitude of the RGB in the LMC are sequence E variables. 
Similarly, \citet{2004AcA....54..347S} and \citet{2007ApJ...660.1486S} 
found 1--2\% (we use 1.5\%) of red giants in the LMC show ellipsoidal 
variations. Moreover, for these fractions, we found the detectability limit 
of the full light curve amplitude is $\sim$0.05 mag for Macho Red band ($M_R$), 
and $\sim$0.025 mag for OGLE $I$ band. It is our aim to use the observed 
fraction of sequence E stars among all the red giants to estimate the 
fraction of PNe with close binary central stars that will be produced by 
a CE event.

\section{Monte Carlo Simulation}\label{sect2}
A Monte Carlo simulation is made to predict the fraction of close binary 
PNe. One million red giant binaries are initially generated by using the 
observed orbital elements distributions. All the binaries are evolved up 
to the RGB and AGB and their evolutionary fates are examined to find which 
stars would produce close binary PNe. The fraction of close binary PNe is 
estimated by using the observed fraction of sequence E stars on the top 1 
magnitude of the RGB.

\subsection{Generating the binary systems}\label{sect2.1}
The simulation requires as input the orbital elements distributions of 
the binary systems. However, these are poorly known in the LMC, so we 
adopt the distributions from binaries in the solar vicinity. Input from 
LMC sources are used when available.

(1) We consider the Initial Mass Function (IMF) as well as the star 
formation history to get the initial mass distribution of the primary 
star. The IMF is assumed to follow \citet{1955ApJ...121..161S}'s power 
law. The LMC star formation history is adopted from \citet{1992ApJ...388..400B}: 
a star burst begins $\sim$4 Gyrs ago and ceases $\sim$0.5 Gyrs ago, with the 
ratio of burst to quiescent star formation rate being 10. According to the 
evolutionary tracks of \citet{2000A&AS..141..371G}, a 4 Gyrs old star has 
a mass of $\rm{1.3~M_{\odot}}$, and a 0.5 Gyrs old star has a mass of 
$\rm{3.0~M_{\odot}}$. The mass range is set from $\rm{0.9~M_{\odot}}$ to 
$\rm{1.85~M_{\odot}}$ for RGB stars, where $\rm{0.9~M_{\odot}}$ is the 
mass for red giants born in the early universe of age 13.7 Gyrs 
\citep{2003ApJS..148..175S}, and $\rm{1.85~M_{\odot}}$ is the upper limiting 
mass for red giants with electron degenerate helium cores on the first giant 
branch. For AGB stars, a full mass range from 0.9 to $\rm{3.0~M_{\odot}}$ is 
allowed.

(2) The mass ratio of the binary systems is drawn from the distribution
of \citet{1991A&A...248..485D}, which follows a Gaussian-type relation,
with its peak at 0.23.

(3) The orbital period of the binary systems is drawn from the distribution of
\citet{1991A&A...248..485D}, which follows a Gaussian-type relation, with 
its peak at 173 years.

(4) We assume the eccentricity is zero. 

(5)The orbital separation is calculated by using the equation of the 
orbital motion. 

(6) The orbital inclination is obtained assuming a random orientation 
of the orbital pole.

\subsection{The properties of the binary systems}\label{sect2.2}
\subsubsection{Mass loss}\label{sect2.2.1}
We consider mass loss from the RGB stars via a stellar wind.
We use the empirical formulation by \citet{1975MSRSL...8..369R} to 
calculate the mass loss rate, but with the rate multiplied by a 
parameter $\eta$ which is set equal to 0.33 
\citep{1983ARA&A..21..271I,2005A&A...441.1117L}.
In order to know the amount of mass lost per magnitude of evolution 
up the red giant branch, we need to calculate the evolution rate.
According to the evolutionary track of \citet{2000A&AS..141..371G}, 
the evolution rate for RGB stars is set to 
d$M_{\rm{bol}}$/d$t$ $\approx$~0.15 mag/Myr.
Due to the mass loss from the primary red giant, the binary system 
loses its orbital angular momentum. The orbital evolution of the 
system is calculated with equation (20) in \citet{2002MNRAS.329..897H}.

AGB mass loss driven by the superwind is not considered explicitly. At 
the AGB tip, stars will lose their mass by a superwind, causing the 
termination of the AGB evolution by ejecting the envelope, just as 
happens for single stars, producing single PNe or wide binary PNe. 
We treat the AGB superwind simply as terminating AGB evolution at
a specific luminosity.

\subsubsection{Stellar radius}\label{sect2.2.2}
For an ellipsoidal binary system containing a red giant, the requirement 
for the detection of the ellipsoidal variability is the minimum fractional 
filling of the Roche lobe. In order to know the minimum filling factor of 
the Roche lobe, we use the light curve generator 
\textsc{nightfall}\footnote{http://www.hs.uni-hamburg.de/DE/Ins/Per/Wichmann/Nightfall.html}
to model the observed ellipsoidal light variations of partial-Roche 
lobe filling systems as a function of binary parameters. The input  
binary parameters are typical of sequence E stars. We find the relation 
between the light curve amplitude ($M_R$) and Roche lobe filling factor `$f$' is:
\begin{equation}
\label{ramp}
\Delta{M_R}=(0.221f^{4}+0.005)\times(1.44956q^{0.25}-0.44956)\times{sin^2i}~~,
\end{equation}
for $0.5<f<0.9,~ 0.1<q<1.5,~0 < i < \frac{\pi}{2}$. We also
find the relation between  $\Delta{M_R}$ and $\Delta{I}$ is well approximated by:
\begin{equation}
\label{iamp}
\Delta{I}=0.87\times{\Delta{M_R}}~~,
\end{equation}
where $\Delta{I}$ is the full light curve amplitude in the $I$ band.

Given the minimum amplitude for detectable light variations (0.05 mag in 
the $M_R$ band or 0.025 in the $I$ band), the minimum radius filling 
factor $f_{\rm min}$ of a red giant in a binary system with detectable 
ellipsoidal light variation can be determined from equations 
(\ref{ramp}) and (\ref{iamp}). With minimum Roche lobe filling factor 
$f_{\rm min}$ , the just-detectable stellar radius is express as:
\begin{equation}
\label{rmin}
R_{\rm{min}}=R_{\rm{L}}\times{f_{\rm min}}~~,
\end{equation}
where $R_{\rm{L}}$ is the equivalent radius of the Roche lobe 
\citep{1983ApJ...268..368E}. When the Roche lobe is filled, the 
maximum stellar radius is:
\begin{equation}
\label{rmax}
R_{\rm{max}}=R_{\rm{L}}~~.
\end{equation}

\subsubsection{Effective temperature and luminosity}\label{sect2.2.3}
The effective temperature and luminosity of the primary red giants are 
calculated separately for O-rich and C-rich stars. To obtain these, we 
use the intermediate age LMC globular cluster NGC 1978 as a template for 
LMC red giant properties.
 
For O-rich stars, the giant branch slope is obtained by using the data
of \citet{2010MNRAS.408..522K}. With the evolutionary tracks of \citet{2000A&AS..141..371G}, 
we make a mass correction for the effective temperature. The zero point of the 
effective temperature $T_{\rm eff} (M_{\rm bol},m)$ is estimated by fitting 
the observed HR diagram of sequence E stars. For C-rich stars, we use the 
giant branch slope obtained from \citet{2010MNRAS.408..522K}. The zero point 
of the $T_{\rm eff} (M_{\rm bol},m)$ relation for C-rich stars is obtained 
by equating the temperatures of the O-rich and C-rich stars at the transition 
luminosity from the O-rich to C-rich stars. The transition luminosity is 
calculated by using the data of \citet{1990ApJ...352...96F} and 
\citet{1994ApJS...92..125V}, with the distant modulus of the LMC equal to 
18.54 \citep{2006ApJ...642..834K}. 

With $L=4\pi \sigma R^{2}T_{\rm{eff}}^{4}$, 
$\rm{log~{\emph{L}/L_{\odot}}=-0.4{(\emph{M}_{bol}-4.75)}}$ and the 
$T_{\rm eff} (M_{\rm bol},m)$ from above, we derive the bolometric 
magnitude for O-rich and C-rich stars as a function of $m$ and $R$. 
If we substitute $R$ with $R_{\rm{min}}$ or $R_{\rm{max}}$, then we 
get the minimum or maximum bolometric magnitude of the red giants with 
just-detectable light variations or with Roche lobes that are just 
full, respectively.

\subsection{Scenario for finding close binary PNe}\label{sect3}
An ellipsoidal variable almost filling its Roche lobe will experience 
a Roche lobe overflow as the star evolves. As a result of the Roche 
lobe overflow, the binary system will suffer a CE event, leading to 
the ejection of the entire envelope. A close binary PN will be produced 
if the primary star is on the AGB, or a post-RGB binary system will be 
formed if the primary star is on the RGB\footnote{We did not consider 
the coalescence of the two companions.}. For RGB binaries, since the 
evolution is too slow after envelope ejection, the system will be unable 
evolve to high effective temperature quickly enough to ionize the ejected 
envelope before it disperses. On the other hand, post-RGB stars will evolve 
slowly through the instability strip where they will be seen as Population 
II Cepheids or RV Tauris. We note that some Population II Cepheids and 
RV Tauris do have luminosities below the RGB tip luminosity 
\citep{1998AJ....115.1921A}. 

\section{Results}\label{sect4}
With the model described in Sect. \ref{sect2}, we calculated the fraction 
of PNNe that are close binaries. The fraction of sequence E stars in the LMC 
was used to normalize the calculation. In order to get the observed fraction 
of sequence E stars, we added single stars until the measured fraction is 
equal to the observed. 

\begin{figure}
\begin{center}
\includegraphics[angle=0,width=0.49\textwidth,height=0.50\hsize]{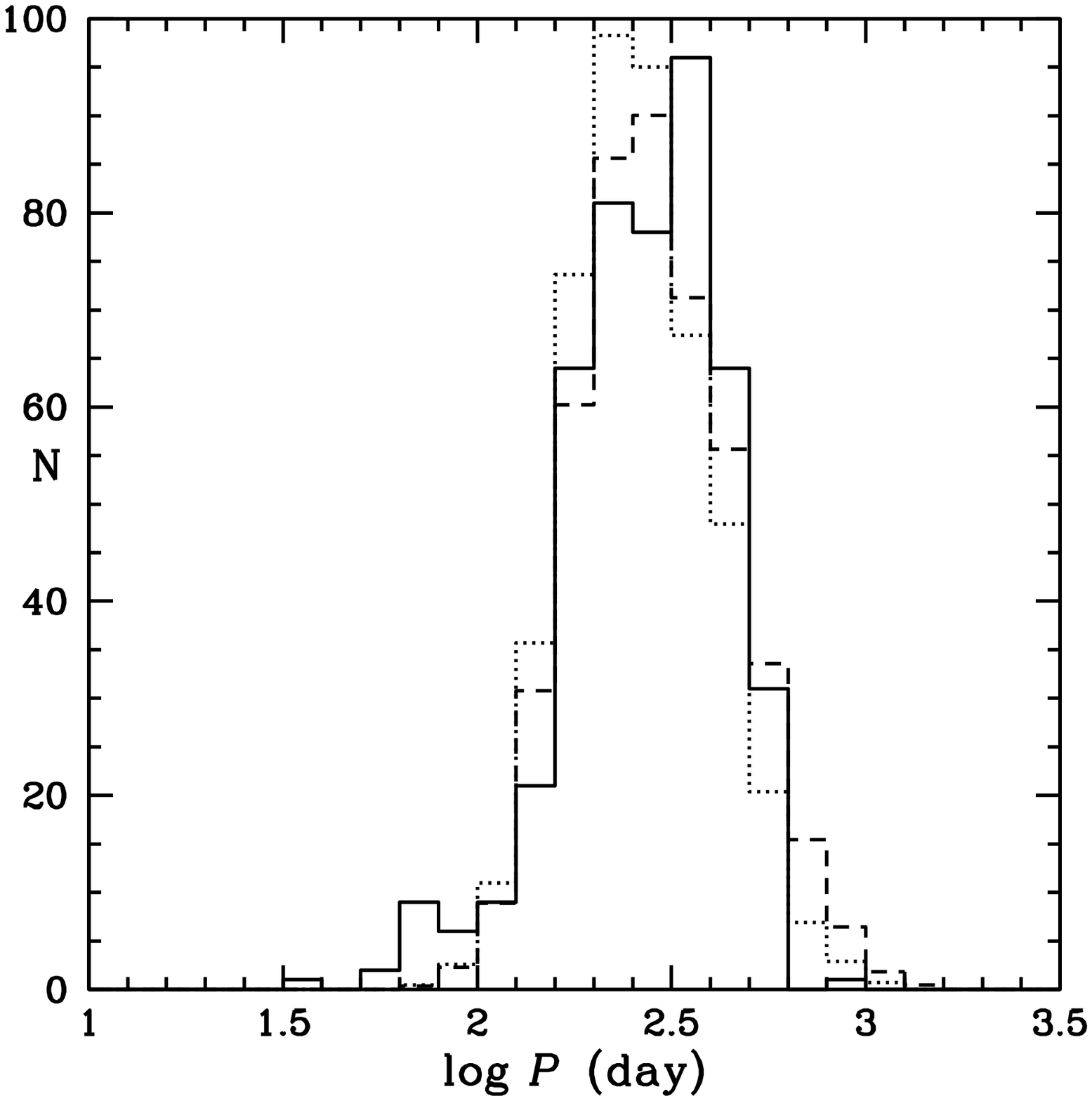}
\includegraphics[angle=0,width=0.49\textwidth,height=0.50\hsize]{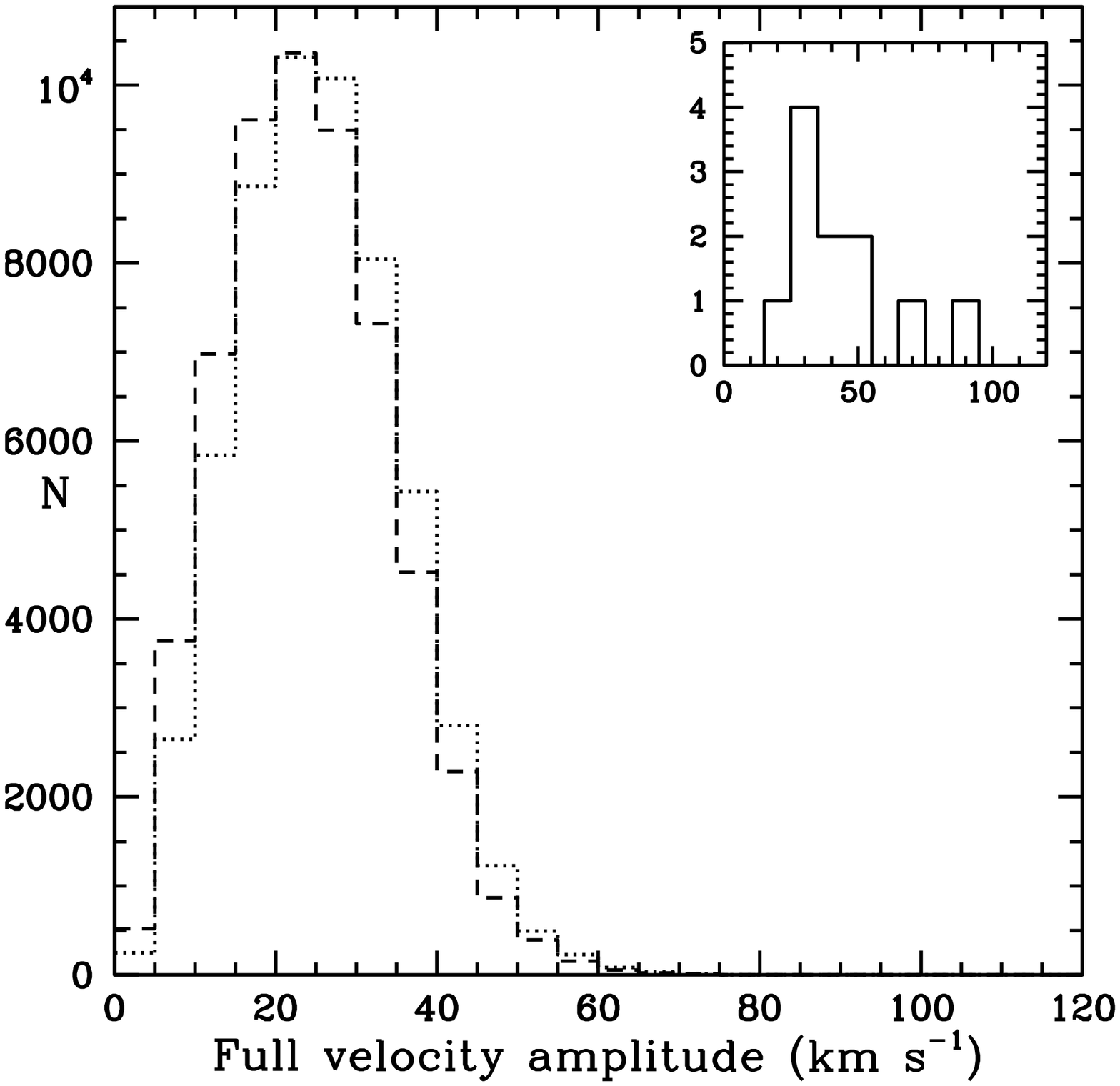}
\caption{The distributions of the orbital period (left) and full velocity 
amplitude (right) of sequence E stars.\label{distribution} Solid line denotes 
distribution from the observation (left: OGLE II; right: \citet{2010MNRAS.405.1770N}),
dotted line denotes distribution from the model by using Wood's frequency, 
short dashed line denotes distribution from the model by using Soszy\'nski's 
frequency.}
\end{center}
\end{figure}

The predicted fraction of close binary PNe is $\sim$10 or 16\%, depending on
whether we use Wood's (0.5\%) or Soszy\'nski's (1.5\%) frequency. 
It indicates that short-period post-CE binaries are a small fraction of 
PNe central stars.

To test the simulation, we predicted the distributions of the period and 
velocity amplitude of sequence E stars and compared them with the observations. 
Fig. \ref{distribution} (left panel) is the period distribution from the 
model and observation (OGLE II). It shows that the simulation reproduces 
the observation well. The period ranges from $\sim$30 to 1000 days, with the 
peak located at $\sim$250 days. We also predicted the full velocity amplitude 
distribution (right panel in Fig. \ref{distribution}). The distribution from 
the model shows a good consistency with the observational results of 
\citet{2010MNRAS.405.1770N} although the observed numbers are small.

An updated and improved version of the results presented here is currently 
being prepared for publication (Nie et al. 2011, in preparation).

\acknowledgements The authors have been partially supported during this
work by Australian Research Council Discovery Project DP1095368.

\bibliography{Nie}

\end{document}